# Tunable Catalysis of Water to Peroxide with Anionic, Cationic, and Neutral Atomic Au, Ag, Pd, Rh, and Os

Kelvin Suggs [1,2], Filmon Kiros[1], Aaron Tesfamichael[2], Zineb Felfli[1] and Alfred Z. Msezane[1]

[1]Department of Physics and Center for Theoretical Studies of Physical Systems, Clark Atlanta University, Atlanta, Georgia 30314, USA

[2]Department of Chemistry, Clark Atlanta University, Atlanta, Georgia 30314, USA

## ABSTRACT

Fundamental anionic, cationic, and neutral atomic metal predictions utilizing density functional theory calculations validate the recent discovery identifying the interplay between Regge resonances and Ramsauer–Townsend minima obtained through complex angular momentum analysis as the fundamental atomic mechanism underlying nanoscale catalysis. Here we investigate the optimization of the catalytic behavior of Au, Ag, Pd, Rh, and Os atomic systems via polarization effects and conclude that anionic atomic systems are optimal and therefore ideal for catalyzing the oxidation of water to peroxide, with anionic Os being the best candidate. The discovery that cationic systems increase the transition energy barrier in the synthesis of peroxide could be important as inhibitors in controlling and regulating catalysis. These findings usher in a fundamental and comprehensive atomic theoretical framework for the generation of tunable catalytic systems.

## Introduction

Catalysis has a wide array of scientific, industrial and biological applications. Industrial operations are often improved by utilizing catalytic technologies and optimization of associated processes. Various chemical reactions such as organic synthesis, combustion, condensation, and biosynthesis are typically utilized to solve practical real-world problems and streamline industrial processes. Research and development is ongoing to further the understanding of reaction mechanisms, theoretical models, and optimal catalytic systems and processes. Specifically, recent investigations utilizing Regge resonances and Ramsauer–Townsend minima obtained through complex angular momentum (CAM) analysis identified negative atomic ions as the fundamental mechanism of catalysis at the atomic scale with the implication that the anionic atomic systems are optimal configurations for catalytic behavior [1]. Moreover, density functional theory calculations have obtained good agreement with the CAM method in determining binding energies and validation of optimal candidates for methane conversion to methanol without $CO_2$ emission [2] and peroxide synthesis [3, 4].

A motivation for further investigation with metallic ions has been inspired by recent studies of graphene intercalated gold molecules, where a tunable band-gap results from the application of an electric field bias perpendicular to the graphene substrate [5]. These results have stimulated the present fundamental study of atomic charge modifications of various transition metals and attendant cationic, neutral, and anionic polarization effects. With these systems we have observed vertical transition state barrier shifts when catalyzing water to peroxide synthesis. This suggests the possibility of implementing tunable catalysis. It is noted that cationic and neutral atomic catalysis of a plethora of systems is a popular approach that is applied to a wide array of systems and has shown great promise. More specifically: 1) Small Au clusters, $Au_2$ to large Au nanoparticles, $Au_{201}$ have been investigated to understand the tunability of the catalytic activity of metal nanoparticles [6]; 2) Temperature-tunability for selective methane catalysis on $Au_2$ has been reported [7]; 3) Highly efficient and broadly applicable cationic Au catalysts are being studied [8]; 4) The search continues for cationic Au complexes for highly efficient catalysts for various processes, including their efficient generation and increased reactivity [9, 10]; 5) Relativistic effects have been investigated to understand the reactivity of Au catalysts [11]; 6) Cationic Pd catalysis has been studied to understand C-H activation [12, 13]; and 7) An organometallic complex has been developed for low temperature oxidation of methane to methanol [14] and ruthenium complexes are being used for the catalytic water oxidation to generate oxygen [15]. Au, Pt and Pd catalysts have been and continue to be developed for the important enyne cycloisomerization process [16, 17]. Furthermore, atomic metals are functionalized typically with polymeric ligands [18-20]. The unifying concept of these functionalization methods is that the electrical inducement of magnetic or dipole properties results in orbital distortion, charge potential modification, and tailored catalytic products.

**Calculation Method**

We have employed the first principles calculations based on Density Functional Theory (DFT) and dispersion corrected DFT approaches for the investigation. For geometry optimization of structural molecular confirmation we utilized the gradient-corrected Perdew-Burke-Ernzerfof (PBE) parameterizations [21] of the exchange-correlation rectified with the dispersion corrections [22]. The double numerical plus polarization basis set was employed as implemented in the DMol3 package [23]. The dispersion-correction method, coupled to suitable density functional, has been demonstrated to account for the long-range dispersion forces with remarkable accuracy. A tolerance of $1 \times 10^{-3}$ Hartrees (Ha) was used with a smearing value of 0.005Ha. Transition state calculations utilized QST algorithms [21] for analysis of energetic barriers. Each calculation consists of 17 atoms including M=1 metallic charged particles, 8 oxygen, and 8 hydrogen atoms where M is Au, Ag, Pd, Rh, and Os. This method has been previously shown to be effective in the investigation of chemical covalent functionalization of perflourophenolazide (PFPA) interaction with graphene [24]. As the calculation of the transition barrier depends crucially on the exchange-correlation scheme employed, the use of reliable dispersion-corrected approach is essential [22]. The error in extracting the transition barrier associated with the transition pathway was estimated to be less than 0.1 eV [24].

**Results and Discussions**

Figure 1(a) presents a transition state barrier path with the approximate height of 1.3eV for peroxide synthesis, void of a catalyst, whereas figure 1(b) approximates an optimal anionic osmium catalyzed peroxide synthesis path reaction whose associated height is 0.15eV. It is important to note that the dramatic reduction of the barrier height from 1.3eV to 0.15eV is indicative of the significant effect of the catalyst on the reaction. In [4] it was demonstrated that the fundamental mechanism of negative ion catalysis in the $H_2O_2$ synthesis using atomic $Au^-$ is the formation of the anionic $Au^-(H_2O)_2$ molecular complex during the transition state, which weakens (breaks) the H-O bonds, thereby promote the formation of the $H_2O_2$ in the presence of $O_2$. The relatively large electron affinity of Au adds to its anionic catalytic effectiveness in the synthesis of $H_2O_2$.

Figure 2(*a*) represents the transition state calculation of anionic osmium catalysis of water to peroxide reaction with molecular orbitals represented by +/- wave functions represented by (transparent) blue and yellow, respectively. Osmium, oxygen, and hydrogen are represented by green, red, and white spheres, respectively. Figure 2(b**)** gives the transition state calculation of anionic osmium catalysis of water to peroxide reaction with molecular orbitals represented by +/- wave functions indicated by (transparent) blue and yellow, respectively. Osmium, oxygen, and hydrogen are represented by green, red, and white spheres, respectively.

Transition state calculations given in figures 3 show the catalyzing effects of cationic, neutral, and anionic Ag, Au, Pd, Rh, and Os atoms when synthesizing water to peroxide. Cationic Au appears to be the least efficient of the given systems as shown by a barrier of ~ 1.3eV. Conversely anionic Os is energetically the most favored with a barrier of ~ 0.15eV as shown by figure 3(b). Analysis of the figure 3(a) shows an ascending energetic barrier ranking order for cationic Au, Pd and Ag, respectively. However, anionic polarization results in a ranking order of Pd, Au, Ag such that Pd and Au switch their relative order and Ag retains its last position among the transition state points.

In figure 3(b) one observes that neutral Rh is superior to neutral Os by about ~ 0.15eV, but the order is reversed when comparing anionic Os and Rh. The anionic barriers exhibit optimal minimum behavior except for anionic Rh with a barrier of approximately 0.45eV. Another interpretation is that ionizing atoms have the analogous effect of inducing a magnetic field on the transition state barrier, resulting in an energetic field gradient tuning effect. Further analysis of charge density plots, Figs. 2(a) and 2(b), reveals charge accumulation on the OH bonds of the transition state barrier. This is a result of hydrogen bond breaking and reformation in the transition state of the system. The atomic ions that lower the transition state barrier the most are the most efficient in hydrogen bond breakage and reformation in the peroxide synthesis process.

Summarizing, it is clear that the anionic Os is the best catalyst for the oxidation of $H_2O$ to $H_2O_2$. It is followed by Pd, Au, Rh and Ag in this order. All the cationic catalysts have a "negative effect" since they produce results that are above the purple line, which represents the standard comparative uncatalyzed reaction energy. For the cationic catalysts Au is the worst, followed by Pd, Ag, Os and Rh, respectively since their effect is negative (increases the barrier). Among the neutral atoms Au is the best catalyst, followed by Ag and Rh, respectively. Both the atomic Pd and Os have a "negative effect" since their results are above the purple line, with the atomic Pd producing a more negative effect compared to Os. It should be noted, however, that neither atomic Au nor atomic Pd catalyzes the $H_2O_2$ reaction as effectively as the anionic Os.

Interestingly, contrary to the anionic all the cationic systems have the effect of increasing the potential energy transition state barriers; their results are above the no catalyst line, viz. the purple line. The order of increase of the energy barrier by the cationic systems is Au, Pd, Ag, Os and Rh. So, a combination of the appropriate systems could provide tunability of catalysis of a plethora of systems, including the oxidation of $H_2O$ to $H_2O_2$. This could obviously be accomplished in some cases through the application of an electric field or through polarization of the atomic systems. Cationic Au is the best for increasing the barrier and is followed by Pd.

## Conclusions

We conclude that anionic atomic systems are optimal and therefore ideal for catalyzing the oxidation of water to peroxide, and that anionic Os is the best candidate among the atomic anions investigated in this work. Furthermore, there is an implication that this result may plausibly extend to relatively negatively charged enzymes in biological systems, polymers, and organic substrates, and poses an interesting hypothesis as to whether this is consistent across interdisciplinary systems, reaction types, atomic cluster substrates and ionic solutions. This conclusion should also be applicable to the oxidation of methane to methanol without $CO_2$ emission [2], with the anionic Os being the best catalyst.

Implementation of transition state barrier approaches to the synthesis of water to peroxide carried out here supports the previous identification of the interplay between Regge resonances and Ramsauer–Townsend minima that characterize the near-threshold electron elastic total cross sections for Au and Pd atoms along with their large electron affinities as the fundamental atomic mechanism underlying nanoscale catalysis [1]. Also important here is the discovery that cationic systems increase the transition energy barrier in the synthesis of peroxide; these could be utilized as inhibitors to control and regulate catalysis.

## Acknowledgements

Research was supported by Atomic, Molecular, and Optical Sciences Program, Chemical Sciences, Geosciences, and Biosciences Division, Office of Basic Energy Sciences, US Department of Energy, and Army Research Office (Grant W911NF-11-1-0194). This research used resources of the National Energy Research Scientific Computing Center, which is supported by the Office of Science of the US DOE under Contract No. DE-AC02–05CH11231. Discussions with Dr. Eric Mintz are appreciated

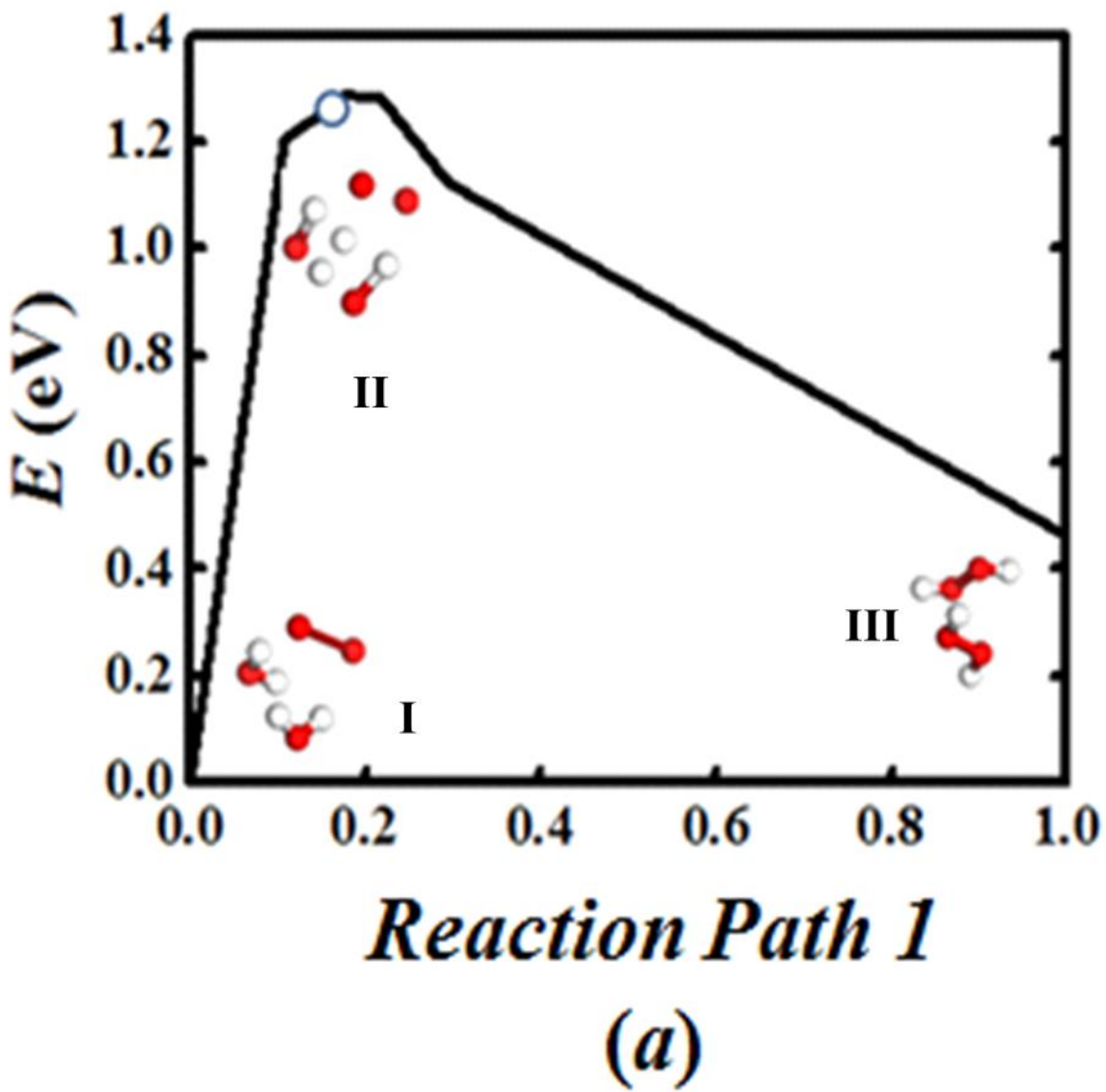

**Figure 1(a):** Transition state potential energy plot of geometrically optimized water to peroxide chemical reaction path 1 with initial (I), transition (II), and final (III) states without the interacting anionic osmium catalyst. Atomic oxygen and hydrogen are represented by red and white spheres, respectively.

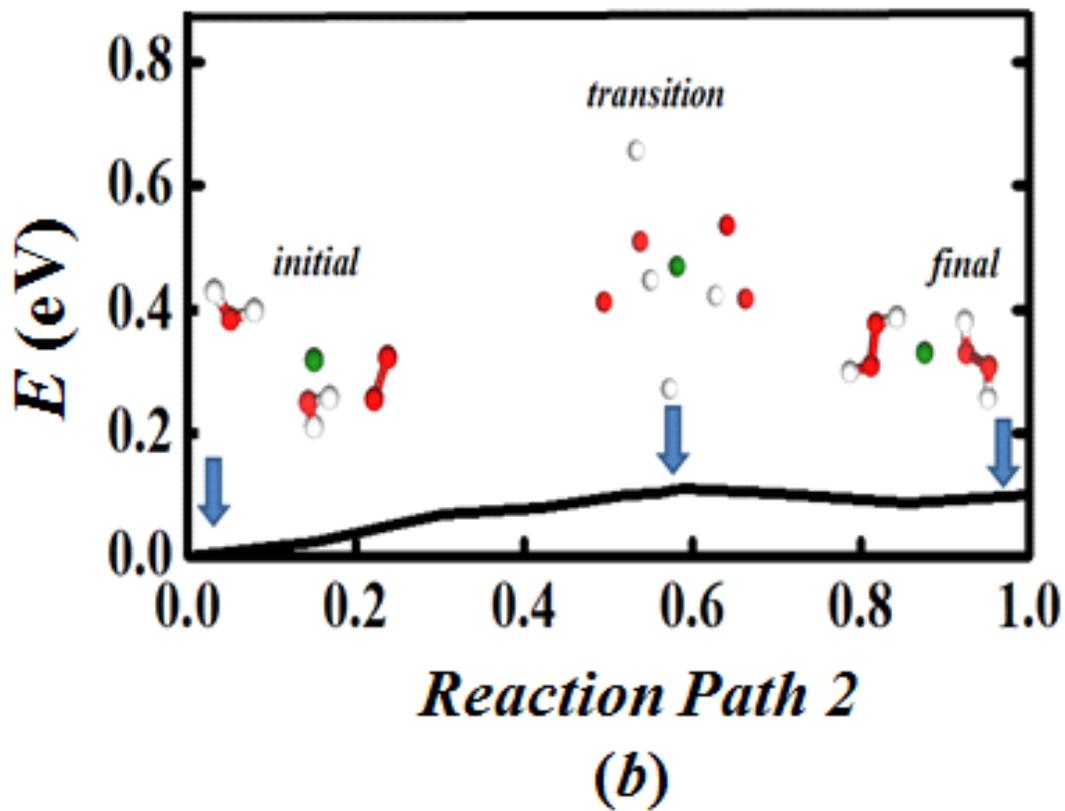

**Figure 1(b):** Transition state potential energy plot of geometrically optimized water to peroxide chemical reaction, path 2 with initial, transition, and final states of interacting anionic osmium catalyst where atomic osmium, oxygen, and hydrogen are represented by olive, red, and white spheres, respectively. Note the dramatic reduction of the transition peak height compared to that of Fig. 1(a)

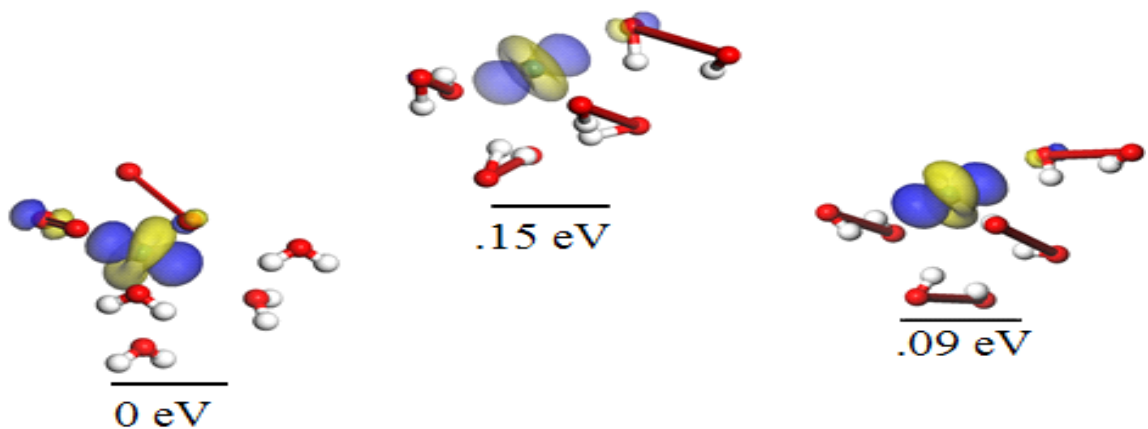

**Figure 2(a):** Transition state calculation of anionic osmium catalysis of water to peroxide reaction with molecular orbitals represented by +/- wave functions indicated by (transparent) blue and yellow, respectively. Osmium, oxygen, and hydrogen are represented by green, red, and white spheres, respectively.

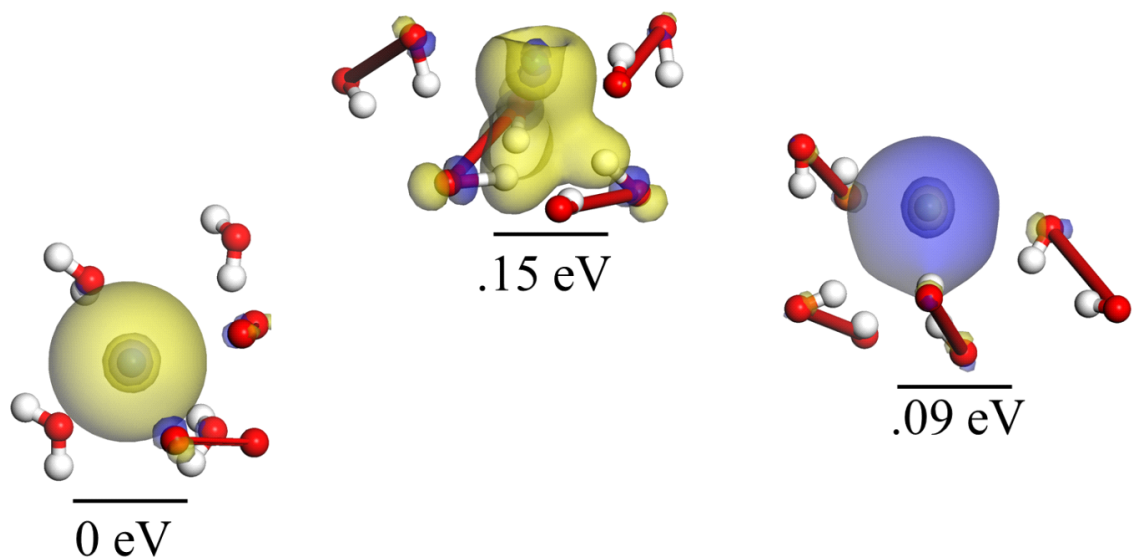

**Figure 2(b):** Transition state calculation of anionic osmium catalysis of water to peroxide reaction with molecular orbitals represented by +/- wave functions indicated by (transparent) blue and yellow, respectively. Osmium, oxygen, and hydrogen are represented by green, red, and white spheres, respectively.

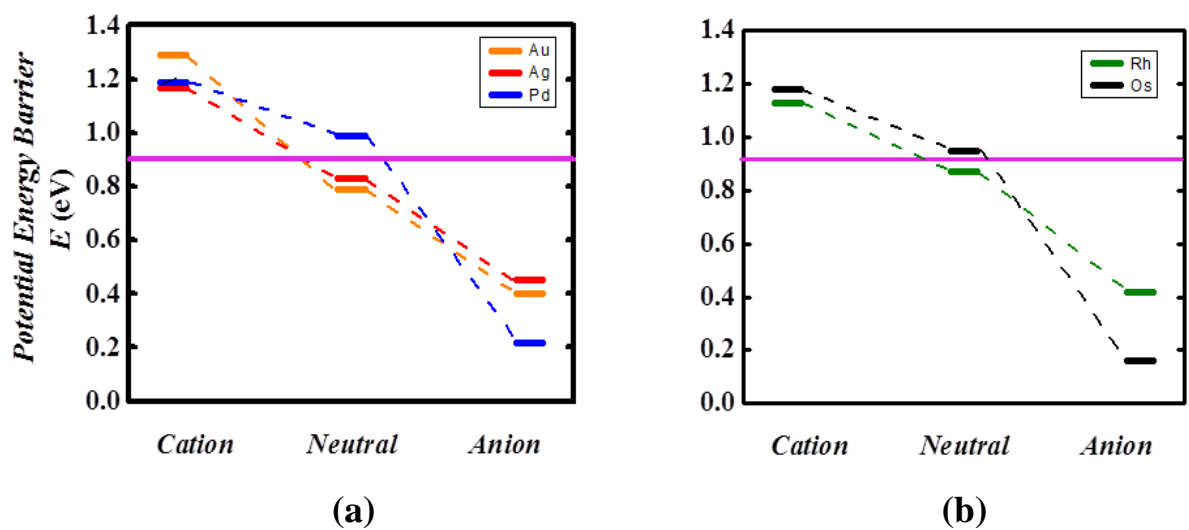

**Figures 3 (a) and (b):** Potential energy transition state barrier calculations of cationic, neutral, and anionic Au, Ag, and Pd, Fig. 3(a) and Rh, and Os, Fig. 3(b) represented by orange, red, and blue and olive, and black spheres, respectively. The standard comparative uncatalyzed reaction energy is represented by the purple line.